\documentclass[pra,aps,amsmath,amssymb,floatfix,twocolumn]{revtex4}
    \usepackage{graphicx}       \usepackage{bm}
    \usepackage{epsf}           \usepackage{epsfig}
\newcommand{\bra}[1]{\mbox{$\left\langle #1 \right|$}}
\newcommand{\ket}[1]{\mbox{$\left|#1\right\rangle$}}
\newcommand{\identity}{\openone}
\begin{document}
\title{State Transfer in Highly Connected Networks and a Quantum Babinet
Principle}

\author{D.~I. Tsomokos$^1$, M.~B. Plenio$^{2,3}$, I. de Vega$^4$ and S.~F. Huelga$^1$}
\affiliation{$^1$Quantum Physics Group, STRI, School of Physics,
Astronomy \& Mathematics, \\University of Hertfordshire, Hatfield
AL10 9AB, United Kingdom} \affiliation{$^2$Institute for
Mathematical Sciences, Imperial College London, London SW7 2PG,
UK} \affiliation{$^3$QOLS, Blackett Laboratory, Imperial College
London, London SW7 2BW, UK} \affiliation{$^4$Max-Planck-Institut
f\"{u}r Quantenoptik, Hans-Kopfermann-Str. 1, Garching D-85748,
Germany}
\date{\today}

\begin{abstract}
The transfer of a quantum state between distant nodes in
two-dimensional networks, is considered. The fidelity of state
transfer is calculated as a function of the number of interactions
in networks that are described by regular graphs. It is shown that
perfect state transfer is achieved in a network of size $N$, whose
structure is that of a $\frac{N}{2}$-cross polytope graph, if $N$
is a multiple of $4$. The result is reminiscent of the Babinet
principle of classical optics. A quantum Babinet principle is
derived, which allows for the identification of complementary
graphs leading to the same fidelity of state transfer, in analogy
with complementary screens providing identical diffraction
patterns.
\end{abstract}
 \maketitle

\section{Introduction}

The dynamics of quantum many-body systems offers a rich variety of
features. This quantum dynamics is often investigated in
one-dimensional chains, which are amenable to exact analytical
treatment in some cases \cite{TakahashiBook} and, in other cases,
efficient numerical methods have been developed for their
simulation \cite{White92}. For more general lattice structures,
henceforth referred to as \emph{graphs}, few analytical treatments
are known. An important problem that arises in this context is the
interplay between the dynamics of quantum many-body systems and
the properties of the underlying graph, which determines the
interaction structure of the many-body system. A variety of
interesting phenomena, two examples of which are perfect state
transfer \cite{njp,ekert_group} and the possibility of deciding
the graph isomorphism problem \cite{AudenaertGRR05}, have recently
been explored in such complex quantum systems.

In the case of graphs with uniform nearest neighbour coupling,
\emph{perfect} state transfer (PST) has so far been proven
possible only with rings of $N=4$ spins, chains of $N=2$ or $N=3$
spins and with Cartesian products of such graphs, the so-called
one-link and two-link hypercubes \cite{ekert_group,facer,simone}.
For larger networks, it appears that increasing the number of
spins and the degree of the underlying graph tends to compromise
the transmission of quantum information \cite{osborne,mauro,kay}.
In the static case it has been shown that higher connectivity and
associated monogamy constraints frustrate the system and affect
its quantum correlations \cite{acin}. On the other hand, if
natural interactions are abandoned in favor of particular coupling
schemes, in which only nearest neighbors interact and the
interaction strength depends on their position relative to a fixed
point, then perfect state transfer is possible in spin chains with
large $N$ \cite{ekert_group, sufiani,nikolopoulos,carlo}.

In this paper we investigate whether it is possible to transfer
perfectly a quantum state between two distant nodes of a
two-dimensional spin network, in which the interactions between
spins are both permanent and homogeneous. We show that PST can be
achieved in such a two-dimensional highly-connected network of
arbitrary size, $N$. This is possible with a unique regular
configuration, namely a two-dimensional graph of the
$\frac{N}{2}$-cross polytope \cite{graphs}, which is dual to the
hypercube in $\frac{N}{2}$ dimensions and isomorphic to a type of
circulant graph \cite{severini}. It turns out that these findings
lead to a natural quantum generalization of a well-known principle
in classical optics. Therefore the plan of the paper is the
following: in Sec. \ref{Sec_2} we introduce a general spin model,
whose defining characteristic is that it preserves the total
number of excitations in the network; then in Sec. \ref{Sec_3} we
present numerical calculations, which reveal the special
properties of $\frac{N}{2}$-cross polytope graphs; and in Sec.
\ref{Sec_4} we provide analytical results that support our
numerical findings and prove the main result of the paper. In Sec.
\ref{Sec_5}, based on the quantum state transfer properties of
complementary graphs, we derive a quantum version of the Babinet
principle from classical optics. Basic results of Monte-Carlo
simulations on the influence of static disorder on the system are
presented in Sec. \ref{Sec_6}. Concluding in Sec. \ref{Sec_7}, we
discuss our results.

\section{Excitation-preserving quantum network \label{Sec_2}}

\begin{figure} \centering
\resizebox{0.80\linewidth}{!} {\includegraphics{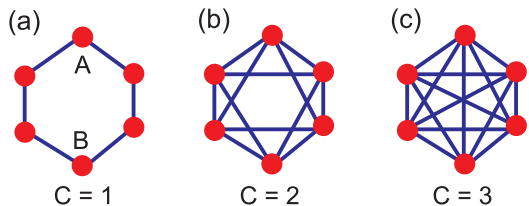}}
\caption{(Color online) A circulant spin network can be used to
transfer quantum states from A to B. In a network of $N=6$ spins
there are three possible configurations with connectivity (a)
$C=1$, (b) $C=2$, and (c) $C=3$, as shown. Network (b) is a
3-cross polytope graph (CPG). \label{Fig_1}}
\end{figure}

We begin by considering $N$ spins-$\frac{1}{2}$ situated along a
circle, as shown in Fig. \ref{Fig_1}. It is understood that if two
spins are interacting, a line is drawn between them. The result is
a graph ${\cal G}=(V,E)$; the vertices $V({\cal G})$ represent the
spin sites and the edges $E({\cal G})$ represent pairwise
interactions. The necessary information about the graph ${\cal G}$
is contained in its adjacency matrix, $A ({\cal G})$, whose
elements are given by $A_{ij} = 1$ if $\{i,j\}\in E({\cal G})$ and
are zero otherwise. We consider Hamiltonians of the form
($\hbar=1$)
\begin{equation}
        {\cal H} = \sum_{k=1}^{N} \omega_k \sigma_k^{+}\sigma_k^{-}
        + \sum_{k\neq l} J_{k,l} (\sigma_k^{-}
        \sigma_{l}^{+} + \sigma_k^{+}\sigma_{l}^{-}),
        \label{NetworkHamiltonian}
\end{equation}
where $\sigma_k^{+}$ ($\sigma_k^{-}$) are the raising and lowering
operators for site $k$, $\omega_k$ is the local site excitation
energy and $J_{k,l}$ denotes the hopping rate of an excitation
between the sites $k$ and $l$. The dynamics in this system
preserves the total excitation number, defined by $\cal
N$=$\sum_{k=1}^N \sigma_k^{+}\sigma_k^{-}$. During dynamical
evolution the state of the network, $\ket{\Psi(t)} = \exp(-i {\cal
H} t) \ket{\Psi_0}$, where $\ket{\Psi_0}$ is the initial state,
always remains in the same excitation sector because $[{\cal
H},{\cal N}]=0$. In what follows we restrict our attention to the
single-excitation sector, for simplicity. In this subspace the
Hamiltonian of the system is equal to the adjacency matrix of the
underlying graph, ${\cal H} = A({\cal G})$, provided that the
spin-spin interactions are homogeneous. Deviations due to
engineering errors in the interactions are also examined later on.
The network is prepared in the state
\begin{eqnarray}
\ket{\Psi_0({\bf j})} \equiv \ket{j} := \ket{0_1 0_2 \cdots 1_j
\cdots 0_N},
\end{eqnarray}
where only spin $j$ is excited. The propagation of an arbitrary
state $\alpha \ket{0_j} + \beta \ket{1_j}$, where $|\alpha|^2
+|\beta|^2 = 1$, is equivalent to the propagation of the state
$\ket{1_j}$ (since the $+1$ eigenstate of $Z_j$, $\ket{0_j}$, does
not evolve under ${\cal H}$). The aim is to transfer the
excitation from $j$ to $N/2 + j$, that is, to the vertex that is
diametrically opposite from $j$ across the ring -- hence we
initially consider that $N$ is even. The state transfer is
quantified by the fidelity
\begin{eqnarray}\label{fidelity}
F(t) := |\bra{\Psi_0({\bf N/2+j})} \exp(-i{\cal H}t)
\ket{\Psi_0({\bf j})}|.
\end{eqnarray}
Perfect state transfer is achieved at a certain time $t_0$ if and
only if $F(t_0)=1$.

\section{Quantum state transfer and connectivity \label{Sec_3}}

\begin{figure}[b] \centering
\resizebox{0.99\linewidth}{!} {\includegraphics{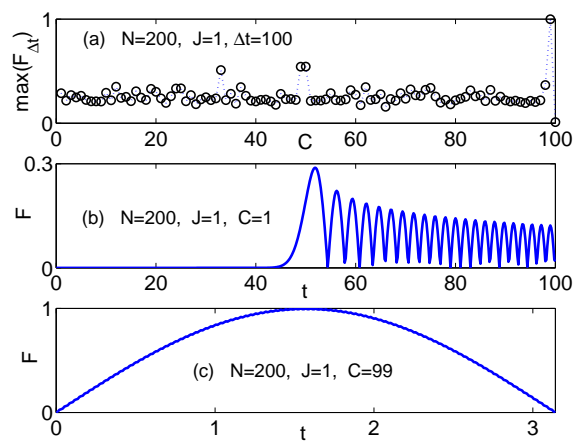}}
\caption{(Color online) (a) Maximum fidelity in the interval
$[0,\Delta t = 100]$ against connectivity for a network of size
$N=200$ and homogeneous interactions. (b) Fidelity against time
for the simple ring network. (c) Fidelity against time for the
$100$-cross polytope graph network. \label{Fig_2}}
\end{figure}

We now ask, `How is the fidelity of quantum state transfer
influenced by the connectivity of a network?' The connectivity,
$C({\cal G})$, is defined here as the number of edges that are
incident on a vertex, counting only within the half-disc defined
by that vertex and the opposite one (i.e., it is half the degree
of the graph). The extreme cases are those of a ring ($C=1$) and a
fully-connected network ($C=N/2$), but in general we have $C =
1,2, \cdots, N/2$ (see Fig. \ref{Fig_1}). Before we analyze this
question analytically we calculate numerically the fidelity $F(t)$
of Eq. (\ref{fidelity}) for $t \in [0,\Delta t]$, given the number
of spins $N$ and the connectivity $C$. The maximum fidelity,
$\max(F_{\Delta t})$, is then determined for the interval $\Delta
t$. It is assumed that $\omega_k = 0$ and $J_{k,l}=1$ in the
Hamiltonian of Eq. (\ref{NetworkHamiltonian}). Therefore the
Hamiltonian of the network is equal to the adjacency matrix of the
underlying graph structure. Under these conditions it is observed
in Fig. \ref{Fig_2}(a) that the fidelity is a non-monotonic and
rather complicated function of the connectivity. However, it
displays remarkable behavior for $C = \frac{N}{2} -1$, which
corresponds to a $2k$-cross polytope graph (CPG) with $N=4k$
spins, where $k$ is a positive integer. In this case, PST is
achieved at $t_0 = \frac{\pi}{2} + n\pi$, i.e., $F(\frac{\pi}{2} +
n\pi)=1$, where $n\ge 0$ is an integer. For $k=1$ we recover the
known result \cite{bose} for a ring with $N=4$.

In Ref. \cite{severini} it was shown that circulant graphs of odd
order \emph{do not allow} perfect state transfer (so our choice of
even $N$ is justified) and, moreover, it was left as an open
question if there exist circulant graphs of even order with $N>4$
that support PST. Our results show that such graphs do indeed
exist: the $2k$-CPG is isomorphic to the circulant graph ${\rm
Ci}_{4k} (1,2,\ldots,2k-1)$. In these networks every spin
interacts with every other spin, except for one (e.g., see Fig.
\ref{Fig_1}(b) for an example). The appropriate choice of $\Delta
t$ is made by comparing trial values with the occurrence time of
the first peak in the evolution of the fidelity for a spin ring
(this evolution is shown in Fig. \ref{Fig_2}(b)). In Fig.
\ref{Fig_2}(c) we show the evolution of the fidelity for a network
with connectivity $C=\frac{N}{2}-1=99$. It is seen that the
fidelity becomes equal to $1$ at $t_0=\pi/2$.

\section{Analytical results \label{Sec_4}}

In this section we analyze the perfect (for $N=4k$) or near
perfect state transfer (for $N=4k+2$) in the configurations
introduced previously. The Hamiltonian $H_{CPG}$ of a cross
polytope graph is that of Eq. (\ref{NetworkHamiltonian}) with
$\omega_k=0$ and $J_{k,l} =
(1-\delta_{l,\frac{N}{2}+k})(1-\delta_{k,l}) = J_{l,k}$. The
Hamiltonian of a fully-connected network, denoted as $H_{fc}$, is
that of Eq. (\ref{NetworkHamiltonian}) with $J_{k,l} =
1-\delta_{k,l} = J_{l,k}$. The Hamiltonian where only opposite
pairs are connected, denoted as $H_{pair}$, has $J_{k,l} =
\delta_{l,\frac{N}{2}+k} = J_{l,k}$ for all $k=1,\ldots,
\frac{N}{2}$.

We start by noting that we have
\begin{eqnarray}
H_{CPG} = H_{fc} - H_{pair}, \;\;\;\;\; [H_{fc},H_{pair}] = 0.
\end{eqnarray}
Therefore, the evolution operator is
\begin{equation}
        e^{-iH_{CPG}t} =  e^{iH_{pair}t} e^{-iH_{fc}t}.
\end{equation}
Due to the fact that
\begin{equation}
H_{pair}=\sum_{k=1}^{N/2} \left(|k\rangle\langle \frac{N}{2}+k| +
|\frac{N}{2}+k\rangle\langle k|\right)
\end{equation}
only couples opposite sites, its dynamics is very simple. It is
straightforward to obtain
\begin{eqnarray*}
        e^{iH_{pair}t} &=&  \identity \cos t + \nonumber \\
        &+& i \sum_{k=1}^{N/2} \left(|k\rangle\langle \frac{N}{2}+k| +
        |\frac{N}{2}+k\rangle\langle k|\right)\sin t.
\end{eqnarray*}
In particular, after a time $t=\frac{\pi}{2}+n\pi$ a starting
state $|k\rangle$ will have been transformed to $(-1)^n
i|\frac{N}{2}+k\rangle$. Finally, to determine the dynamics of
$H_{CPG}$ we need to consider $H_{fc}$. The latter can be
expressed as
\begin{equation}
H_{fc} = N|+\rangle\langle +| - \identity,
\end{equation}
where
\begin{equation}
|+\rangle = \frac{1}{\sqrt{N}}\sum_{k=1}^N |k\rangle.
\end{equation}
We have
\begin{equation*}
        e^{-iH_{fc}t} = \left[|+\rangle\langle +| e^{-iNt}
        + (\identity -|+\rangle\langle +|)\right]e^{it}.
\end{equation*}
Therefore, a state $|k\rangle$ is mapped onto itself, up to a
global phase, under $e^{-iH_{fc}t}$ when $Nt=2\pi k$ with
$k\in\mathbb{N}$.

As a consequence, the dynamics due to $H_{CPG}$ allows for PST if
both $Nt=2\pi k$ and $t=\frac{\pi}{2}$ are satisfied for the same
$t$. This implies the condition
\begin{equation}
        N = 4k
\end{equation}
and explains the possibility of PST in $2k$-cross polytopes. For
$N=4k+2$ the analysis above immediately applies and shows that we
do not have PST at $t=\frac{\pi}{2}$.

More generally, we can find the transfer fidelity for
$t=\frac{\pi}{2}$. Starting with $|k\rangle$ and using
$e^{iH_{pair}t}|+\rangle = e^{it}|+\rangle$ we find at
$t=\frac{\pi}{2}$ the state
\begin{eqnarray*}
        e^{iH_{pair}t} e^{-iH_{fc}t} \ket{+} &=&\\
        && \hspace*{-3.cm} =-\left[\frac{1}{\sqrt{N}} \ket{+} e^{-iN\pi/2}
        + \left(\ket{\frac{N}{2}+ k} - \frac{1}{\sqrt{N}}\ket{+} \right)
        \right].
\end{eqnarray*}
Then the fidelity, $|\bra{\frac{N}{2}+k} e^{iH_{pair}t}
e^{-iH_{fc}t} \ket{k}|^2$,  is
\begin{eqnarray}
        F = 1 - \frac{2}{N} \left(1-\frac{1}{N}\right) \left[1-\cos\frac{N\pi}{2}\right].
\end{eqnarray}
For $N=4k$ we recover $F=1$, while for $N=4k+2$ we find that
$F=(1-2/N)^2$. Therefore, as $N\rightarrow \infty$, the fidelity
approaches $1$ and we obtain almost PST.

\section{Quantum Babinet Principle \label{Sec_5}}

\begin{figure}[t] \centering
\resizebox{0.90\linewidth}{!} {\includegraphics{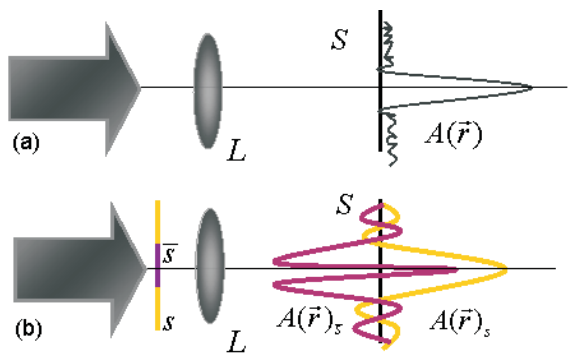}}
\caption{(Color online) Illustration of Babinet's principle in an
optical set up with Fraunhofer conditions. (a) An unobstructed
plane wave is focused by a lens $L$ and produces a diffraction
pattern of amplitude $A(\vec{r})$ on the screen $S$. (b)
Diffraction patterns resulting from complementary screens $s$ and
$\bar{s}$, whose opaque and transparent areas are swapped. At any
point downstream from $s$ and $\bar{s}$, the sum of the two
diffracted amplitudes, $A_s(\vec{r})+A_{\bar{s}}(\vec{r})$, equals
the amplitude diffracted from the unobstructed plane wave
$A(\vec{r})$. Away from the central spike, this amplitude is zero
and therefore $A_s(\vec{r})=-A_{\bar{s}}(\vec{r})$, which leads to
Babinet's prediction of identical diffracted light fields for
complementary apertures. Complementary apertures play the role of
complementary graphs describing quantum spin networks. Increasing
$\vec{r}$ corresponds to increasing the number of nodes
$N$.\label{Fig_3}}
\end{figure}

These results provide a clear insight into the basic mechanisms
that facilitate PST in these systems. The key realization is that
a fully connected network in which some couplings $J_{k,l}$ are
removed, can behave similarly to an initially unconnected network
which is supplemented with the very same $J_{k,l}$ links. This
result is in fact reminiscent of the Babinet principle of
classical optics \cite{babinet}, which is illustrated in Fig.
\ref{Fig_3}. In our context of state transfer through connected
networks, the situation is similar in the sense that
\begin{equation*}
e^{-iH_{CPG}t}e^{-iH_{pair}t}= e^{-iH_{fc}t}
\end{equation*}
because $H_{CPG}$ and $H_{pair}$ commute and also $e^{-iH_{fc}t}$
equals the identity at specific times $t$ (in the optical setting
this is the situation when all incident light emerges unaffected).
Of course in the quantum setting we have the added problem that
$e^{-iH_{CPG}t}e^{-iH_{pair}t}\neq e^{-i(H_{CPG}+H_{pair})t}$ in
general.

The analog of the Babinet theorem does hold however for much more
general settings than just that of commuting $H_{CPG}$ and
$H_{pair}$. Indeed, as before, let us assume that
\begin{eqnarray}
H_f = N\mathbb{P}, \;\;\;\;\; \mathbb{P} = |+\rangle\langle +|.
\end{eqnarray}
For a sequence of $H_N$ that satisfies
\begin{eqnarray}
||\mathbb{P}H_N(\openone-\mathbb{P})+(\openone-\mathbb{P})H_N\mathbb{P}||
= O(\frac{1}{\sqrt{N}})
\end{eqnarray}
we compare the dynamics of $H_N$ and $H_f-H_N\equiv H_c$ in the
limit of large $N$. The following argument is not fully rigorous
in that it does not provide detailed error estimates but these may
be provided in a more detailed analysis.

Let us consider the dynamics under $H_c$ in an interaction picture
with respect to $H_f$ when this Hamiltonian becomes
time-dependent,
\begin{equation*}
H_{c,I}(t)=e^{iH_f t}(H_c-H_f)e^{-iH_f t} = -e^{iH_f t}H_N
e^{-iH_f t}.
\end{equation*}
The corresponding time-evolution operator from $t_1$ until $t_2$
in the interaction picture will be denoted by $U_I(t_1\rightarrow
t_2)$. Now we note that $H_{c,I}(t)$ may contain rapidly
oscillating terms (those coupling subspace defined by $\mathbb{P}$
to the subspace defined by $\openone - \mathbb{P}$) thanks to the
action of $e^{iH_f t}$. These rapidly oscillating terms may be
neglected for large $N$ leading to a correction of order $1/N$ in
the dynamics. Hence, we find
\begin{eqnarray}
        H_{c,I}(t) &\cong& -(\openone - \mathbb{P})H_N(\openone - \mathbb{P})
        - \mathbb{P} H_N \mathbb{P}\\
        &=& -H + \mathbb{P}H_N(\openone - \mathbb{P})+(\openone -
        \mathbb{P})H_N\mathbb{P}. \nonumber
\end{eqnarray}
As we had assumed earlier that
$||\mathbb{P}H_N(\openone-\mathbb{P})+(\openone-\mathbb{P})
H_N\mathbb{P}||$ is of order $1/\sqrt{N}$ we find that
$H_{c,I}(t)$ is well approximated by $-H_N$ up to corrections that
decrease with increasing $N$. Hence $U_I(0\rightarrow t)\approx
e^{iH_N t}$, and we find that
\begin{equation}
e^{-iH_ct} = e^{-iH_f t} U_I(0\rightarrow t) \cong e^{-iH_f
t}e^{iH_N t}.
\end{equation}
Now we consider the transfer fidelity from state $|k\rangle$ to
$\ket{\frac{N}{2} + k}$, as an example. The amplitude $\bra{
\frac{N}{2}+k} e^{-iH_ct} \ket{k}$, using $e^{-iH_ct} =
e^{-iH_ft}e^{iH_Nt}$, is found to be equal to
\begin{eqnarray*}
        && \bra{\frac{N}{2}+k } e^{iH_N t} \ket{k}
        + \langle \frac{N}{2}+k|+\rangle \langle +|e^{iH_Nt}|k\rangle
        (e^{iNt}-1) \\
        && \cong \langle \frac{N}{2}+k|e^{iH_Nt}|k\rangle
\end{eqnarray*}
where the difference decreases with increasing $N$. Therefore, the
transition amplitudes according to the dynamics under $-H_N$ and $H_c$
are asymptotically (in $N$) equal. Note that for a real Hamiltonian
$H_N$ we have $(\langle \frac{N}{2}+k|e^{iH_Nt}|k\rangle)^*=\langle
\frac{N}{2}+k|e^{-iH_Nt}|k\rangle$ so that
\begin{eqnarray}
        |\langle \frac{N}{2}+k|e^{-iH_ct}|k\rangle| &\cong&
        |\langle \frac{N}{2}+k|e^{-iH_Nt}|k\rangle|
\end{eqnarray}
again with an error that decreases with increasing $N$. This
is the quantum Babinet principle.

\section{Influence of disorder \label{Sec_6}}

We provide here a brief analysis of realistic engineering errors
in the interactions of a $2k$-CPG network in order to assess the
robustness of a possible experimental implementation. We take into
account two types of errors: (i) disorder in the interactions, and
(ii) random breaking of interactions. For case (i) we assume that
if $p$ and $q$ are interacting then the interaction strength can
take any value in the interval $[1-\delta, 1+\delta]$, with equal
probability. The amount of disorder is thus quantified by $\delta
\in [0,1]$. In case (ii) some interactions are randomly broken,
that is, $J_{pq}$ vanishes for a fixed number of pairs $(p,q)$.
The number of broken interactions is $B\in [0,1)$, given as a
ratio to the total number of interactions in the network. The main
results of Monte-Carlo simulations on $\frac{N}{2}$-CPG networks
with $N=40,80,120,200,400$ spins, are as follows. For type-(i)
errors we find that disorder up to $\delta = 0.02$ allows for
almost PST in smaller networks ($N < 100$). In particular, the
maximum fidelity $F$ is greater than $0.99$, on average, with a
worst-case value of $0.98$ in the case of $N=40$; while for
$N>100$ the average maximum $F$ is over $0.95$ for disorder that
is less than $2\%$. For type-(ii) errors we find that the random
breaking of very few bonds, so that $B<0.001$, still allows for
very high quality state transfer, where the maximum $F$ is larger
than $0.95$, on average. However, the value of the worst-case
fidelity peak fluctuates considerably on individual cases,
depending on the positions of the broken bonds.

In this connection, the usefulness of the quantum Babinet
principle can be illustrated in the case of transport of
excitations through noisy networks, a setting that has recently
been introduced independently in \cite{AspuruGuzik08} and
\cite{PlenioH08}. Initially all population resides in a given site
and we evaluate how much population may be transferred
asymptotically to a selected target site. To this end, we let the
target site be attached to a sink to which the population is
transferred irreversibly. We want to analyze whether the presence
of local dephasing can assist the excitation transfer. If the sink
is attached to site $\frac{N}{2}+1$, then the Babinet principle
implies that the evolution is that of a system where only the
opposite two sites are coupled, and we recover a situation for
which it was proven in \cite{PlenioH08} that no dephasing enhanced
transport is possible \cite{new}.

\section{Summary and Discussion \label{Sec_7}}

We have shown that PST is achieved in a network of size $N$, whose
structure is that of a $\frac{N}{2}$-cross polytope graph, if $N$
is a multiple of $4$. If $N$ is even, but not a multiple of $4$,
then almost PST is achieved for larger networks of this kind, so
that $F$ approaches 1 for $N\rightarrow \infty$. These results can
be interpreted in terms of a quantum Babinet principle, which
establishes the conditions required for having {\em complementary}
graphs leading to same fidelity of state transfer, in analogy with
the classical situation of obtaining identical diffraction
patterns from complementary screens. As shown in various examples,
invoking Babinet's principle alone can simplify the analysis of
the performance of connected networks and therefore become a
useful tool in tackling a variety of problems in quantum
information theory.

\section*{Acknowledgements}
This work was supported by the EU via Integrated Projects QAP and
SCALA, STREP actions CORNER and HIP and the EPSRC through the
QIP-IRC. DIT acknowledges the EPSRC for support in the form of a
research fellowship (EP/D065305/1). MBP holds a Royal Society
Wolfson Research Merit Award. IDV acknowledges support from
Ministerio de Educacion y Ciencia, Spain. We thank Neil Oxtoby for
careful reading of the manuscript.

\emph{Note added.---} Upon completion of the present work we
became aware of the closely related work of Ref.
\cite{bose_recent}, in which a similar result is established using
different methodology. We would like to thank Simone Severini for
useful correspondence.



\begin{thebibliography}{99}

\bibitem{TakahashiBook}
M. Takahashi, {\em Thermodynamics of One-Dimensional Solvable
Models}, Cambridge University Press, 2005.
%
\bibitem{White92} S.~R. White, Phys. Rev. Lett. {\bf 69}, 2863 (1992).
%
\bibitem{njp}
M.~B. Plenio, J. Hartley and J. Eisert, New. J. Phys. {\bf 6}, 36
(2004).
%
\bibitem{ekert_group}
M. Christandl, N. Datta, A. Ekert and A.~J. Landahl, Phys. Rev.
Lett. \textbf{92}, 187902 (2004);  M. Christandl, N. Datta, T.~C.
Dorlas, A. Ekert, A. Kay, and A.~J. Landahl, Phys. Rev. A {\bf 71},
032312 (2005).
%
\bibitem{AudenaertGRR05} K. Audenaert, Ch. Godsil, G. Royle and
T. Rudolph, Journal of Combinatorial Theory {\bf 97}, 74 (2007);
E-print arXiv:math/0507251.
%
\bibitem{facer}
C. Facer, J. Twamley, J. Cresser, Phys. Rev. A \textbf{77}, 012334
(2008).
%
\bibitem{simone}
A. Bernasconi, C. Godsil, S. Severini, `Quantum Networks on
Cubelike Graphs', arXiv:0808.0510v1.
%
\bibitem{osborne}
T.~J. Osborne and N. Linden, Phys. Rev. A \textbf{69}, 052315
(2004).
%
\bibitem{mauro}
M. Paternostro, G.~M. Palma, M.~S. Kim and G. Falci, Phys. Rev. A
\textbf{71}, 042311 (2005).
%
\bibitem{kay}
A. Kay, Phys. Rev. A \textbf{73}, 032306 (2006).
%
\bibitem{acin} A. Ferraro, A. Garc\'ia-Saez and A. Ac\'in,
Phys. Rev. A \textbf{76}, 052321 (2007).
%
\bibitem{nikolopoulos}
V. Ko\v{s}t\'{a}k, G.~M. Nikolopoulos, I. Jex, Phys. Rev. A
\textbf{75}, 042319 (2007).
%
\bibitem{sufiani}
M.~A. Jafarizadeh and R. Sufiani, Phys. Rev. A \textbf{77}, 022315
(2008).
%
\bibitem{carlo}
C. Di Franco, M. Paternostro, D.~I. Tsomokos, S.~F. Huelga, Phys.
Rev. A \textbf{77}, 062337 (2008).
%
\bibitem{graphs}
H.~S.~M. Coxeter, \emph{Regular Polytopes}, 3rd ed., New York:
Dover Publications (1973).
%
\bibitem{severini}
N. Saxena, S. Severini, I.~E. Shparlinski, Int. J. Quant. Inf.
\textbf{5}, 417 (2007).
%
\bibitem{bose}
M-H. Yung and S. Bose, Phys. Rev. A {\bf 71}, 032310 (2005).
%
\bibitem{babinet}
See M. Babinet, C. R. Acad. Sci. {\bf 4}, 638 (1837) for the
original reference. This result is featured in most classical
optics textbooks, for instance, see pp. 49-50 of G. Brooker
\emph{Modern Classical Optics (Oxford Master Series in Physics)},
Oxford University Press, 2003.
%
\bibitem{AspuruGuzik08} M. Mohseni, P. Rebentrost, S. Lloyd
and A. Aspuru-Guzik, arXiv:0805.2741; P. Rebentrost, M. Mohseni,
and A. Aspuru-Guzik, arXiv:0806.4725;
P. Rebentrost, M. Mohseni, I. Kassal, S. Lloyd, A. Aspuru-Guzik,
arXiv:0807.0929.
%
\bibitem{PlenioH08} M.~B. Plenio and S.~F. Huelga, arXiv:0807.4902.
%
\bibitem{new}
More general settings that allow for dephasing assisted transport
will be presented elsewhere.
%
\bibitem{bose_recent}
S. Bose, A. Casaccino, S. Mancini, S. Severini, arXiv:0808.0748v1
(2008), `Communication in XYZ All-to-All Quantum Networks with a
Missing Link'.
%
\end{thebibliography}
\end{document}